\documentclass[aps,prl,superscriptaddress,twocolumn,showpacs,longbibliography,amsmath,amssymb,nofootinbib,floatfix]{revtex4-2}
\usepackage[utf8]{inputenc}
\usepackage[T1]{fontenc}
\usepackage{microtype}
\usepackage{graphicx}
\usepackage{mathbbol}
\usepackage{amssymb}
\usepackage{dsfont}
\usepackage{comment}
\usepackage[usenames,dvipsnames,svgnames,table]{xcolor}
\usepackage[colorlinks,linkcolor=blue,citecolor=blue,urlcolor=blue]{hyperref}
\usepackage{siunitx}

\newcommand\downtriangle{\rotatebox[origin=c]{180}{\scriptsize$\triangle$}}

\def\equationautorefname~#1\null{Eq.~(#1)\null}

\newcommand{\braket}[1]{\ensuremath{\langle{#1}\rangle}}
\newcommand{\bra}[1]{\langle #1 |}
\newcommand{\ket}[1]{| #1 \rangle}

\def\equationautorefname~#1\null{Eq.~(#1)\null}

\begin{document}
	\title{Dynamical gauge fields with bosonic codes}
	\date{\today}
	\author{Javier del Pino}
	\affiliation{Institute for Theoretical Physics, ETH Zürich, 8093 Zürich, Switzerland}
	\author{Oded Zilberberg}
	\affiliation{Department of Physics, University of Konstanz, 78464 Konstanz, Germany}
	
	\DeclareGraphicsExtensions{.pdf,.png,.jpg}
	
\begin{abstract}
	    The quantum simulation of dynamical gauge field theories offers the opportunity to study complex high-energy physics with controllable low-energy devices. For quantum computation, bosonic codes promise robust error correction that exploits multi-particle redundancy in bosons. Here, we demonstrate how bosonic codes can be used to simulate dynamical gauge fields. We encode both matter and dynamical gauge fields in a network of resonators that are coupled via three-wave-mixing. The mapping to a $\mathbb{Z}_2$ dynamical lattice gauge theory is established when the gauge resonators operate as Schrödinger cat states. We explore the optimal conditions under which the system preserves the required gauge symmetries. Our findings promote realising high-energy models using bosonic codes. 
\end{abstract}

\maketitle	
Dynamical gauge field theories (DGFTs), commonly appearing in high-energy physics, are extremely challenging to simulate or solve using classical methods. A prototypical example is quantum electrodynamics, a theory that features `matter' degrees of freedom interacting in the presence of  `light' gauge fields. The two constituents experience complex combined dynamics, e.g., via the change in the electric field caused by the redistribution of charges. The gauge fields in electrodynamics are Abelian, i.e., their quantum operators commute. In this `simple' DGFT, we have a continuous $U(1)$ degree of freedom with evolution in an infinite-dimensional Hilbert space. In other realms, even more complex DGFTs appear, e.g., in quantum chromodynamics, where they describe quark-gluon interactions in heavy-ion collisions, or the deep core of neutron stars~\cite{Wilson1974, Gattringer2010}.

In recent years, quantum simulators have become a reality, leading to the emulation of numerous effects, such as topological phases of matter~\cite{kraus2012topological,verbin2013observation,lohse2016thouless,lohse2018exploring,zilberberg2018photonic}, collective phenomena~\cite{carusotto2013quantum,gross2017quantum,ferri2021emerging}, cascade effects~\cite{schreiber2015observation,goblot2020emergence}, and many more. Such technologies also offer the means to controllably realise DGFTs~\cite{Wiese2013, Banuls2020, Zohar2016, Altman2021}. The simulation relies on mapping DGFTs to simplified lattice gauge theories~\cite{Kogut1975} (LGTs). This map involves confining the matter degrees of freedom to move only between nodes of a lattice, while the gauge fields are associated with the links between the nodes~\cite{Kogut1975,suppmat}. Further simplifications involve using low-dimensional `quantum links' that replace infinite-dimensional gauge fields, while preserving the symmetries of the original model~\cite{chandrasekharan1997quantum, Wiese2013, Dutta2017}. In addition, minimal $\mathbb{Z}_2$-symmetric LGTs appeared and have sparked interest in spin~\cite{Kogut1975}, fermionic~\cite{Borla2020} and bosonic  models~\cite{Gonzalez-Cuadra2020}, as well as the toric code for quantum error correction~\cite{Kitaev2003}.

The quantum simulation of DGFTs, however, is very challenging. Most proposals thus far harness `static' gauge fields, which act on the matter without reacting, and strive to make them dynamic~\cite{Goldman2014b, Goldman2014c, Walter2016b, Zapletal2019b}. The static fields are an established instrument for breaking spatio-temporal symmetries used to induce nontrivial topological phenomena in lattice systems~\cite{Umucalilar2011, Fang2012, Peano2015, Ozawa2019,mathew2018synthetic,delpino2022}. Rendering gauge fields dynamic, LGTs simulators are being developed in various quantum simulation technologies~\cite{Hauke2013, Monroe2019, Lewenstein2007, PhysRevLett.109.175302, Wiese2013, Zohar2016, Barbiero2019, Surace2020, Ballantine2017,Marcos2013, Marcos2014, Homeier2021, Martinez2016, Gorg2019, schweizer2019floquet,Yang2020,doi:10.1126/science.abl6277}. Yet, despite the variety of proposals, the constraints that LGTs must fulfil impede a clear experimental realisation. 

\begin{figure}
\includegraphics[width=\linewidth]{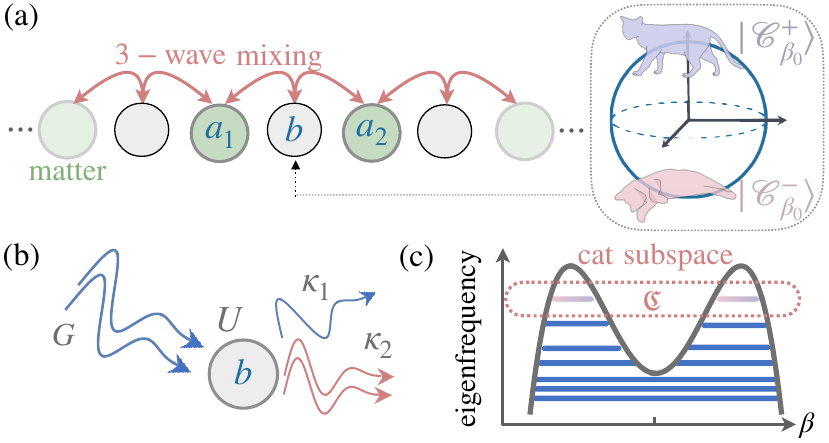}
\caption{(a) Chain consisting of resonators of type $a$ and  $b$ labelled as matter and gauge modes, respectively [cf.~Eq.~\eqref{eq:full_H}]. The resonators are coupled via three-wave mixing interactions [cf.~Eq.~\eqref{eq:3WM}]. (right zoomed box) We show that the system can simulate a 1+1 D LGT when the gauge resonators are projected onto discrete bosonic code subspaces, namely, into cat states $\ket{\mathcal{C}^\pm_{\beta_0}}$ with parity $\pm$ (see text). (b) Sketch of the `link' site as a Kerr parametric resonator [cf.~Eq.~\eqref{eq:KerrParam}] subject to single- ($\kappa_1$) and two-photon ($\kappa_2$) loss. (c) The interplay between the drive $G$ and the nonlinearity $U$ stabilises nonclassical states, e.g., cat states that appear as ground states of the Hamiltonian~\eqref{eq:KerrParam}, with amplitudes $\beta_0$ that minimise the mean-field double-well potential $\langle H_{\rm field}\rangle_{\rm MF}=-U|\beta|^{4}+G\mathrm{Re}(\beta^{2})$.} \label{fig:1}
\end{figure}

Recently, bosonic error correction codes have spurred much interest as a possible route for robust and scalable quantum computing~\cite{Chuang1997, Terhal2020, Joshi2020, Cai2021,girvin2021introduction}. These codes exploit the infinite-dimensional Hilbert space of a harmonic oscillator for redundant information encoding, thus significantly reducing hardware requirements~\cite{suppmat}. Remarkably, the realisation of the so-called `cat-qubit', the workhorse of bosonic codes~\cite{Leghtas2013}, has been recently proposed and demonstrated in nonlinear microwave cavities~\cite{Vlastakis2013,Wang2016, Puri2019, grimm2019kerr}. A wide variety of platforms could potentially allow for bosonic `cat codes', including cavity-QED~\cite{Hacker2019}, trapped ions~\cite{Wineland2013}, bulk acoustic-wave resonators~\cite{bild2022schr}, nanomechanical~\cite{radic2022nanomechanical}, and  optomechanical~\cite{Hoff2016} systems. The prospects for error correction make bosonic error correction platforms an interesting candidate also for quantum simulation~\cite{suppmat}.

In this work, we demonstrate how a $\mathbb{Z}_2$ LGT can be realised using a bosonic code setup. Specifically, we consider a coupled resonator network whose nodes embody both matter and gauge fields. The gauge resonators are nonlinear and driven into `Schrödinger cat' states that encode the dynamical gauge fields. Three-wave-mixing between the network resonators engenders a gauge-invariant light-matter coupling. We explore the bounds for the ideal operation of our scheme and show that it requires the cat states to have a finite coherent amplitude and the three-wave mixing to be moderate. In this regime, we obtain similar functionality and scaling strategies as contemporarily used in standard-qubit-based LGTs~\cite{Banuls2020,Altman2021}. Concurrently, we benefit from the substantial advantage of bosonic codes compared with qubit-based approaches, e.g., their reduced hardware overhead due to efficient quantum-error correction. Furthermore, a compact simulator assembled using our scheme can be realised with various platforms, including photons, phonons, magnons, and polaritons.

We consider a 1D chain of coupled resonators, see Fig.~\ref{fig:1}(a). The resonators are alternatively positioned with matter nodes and gauge sites. The latter will act as links in the following. The chain's building block is a matter-field link, [labelled sites in Fig.~\ref{fig:1}(a)] governed by the Hamiltonian
\begin{align}\label{eq:full_H}
H=H_\text{matter}+H_\text{field}+H_\text{coup}\,.
\end{align}
The matter nodes' Hamiltonian reads
\begin{align}
H_\text{matter} =& \sum_{i={1,2}} \omega_i a_i^{\dagger}a_i^{\phantom\dagger}\,,
\end{align}
with bosonic annihilation operators $a_i$ ($i=\{1,2\}$) and frequency $\omega_i$. 
The link's gauge site is also a resonator [cf.~Eq.~\eqref{eq:KerrParam} below] with bosonic annihilation operator $b$, and frequency $\omega_b$. Gauge resonators are operated in the bosonic code regime. Specifically, we assume that they operate as Schrödinger cat states, i.e., that their state is spanned by 
$\ket{\mathcal{C}_{\beta}^\pm}\sim(\ket{\beta}\pm\ket{-\beta})$, where $\ket{\beta}$ denotes a coherent state with amplitude $\beta$~\cite{Cochrane1999, Wielinga1993}. The $\pm$ label even/odd photon number parities, where the parity operator reads $\Pi(n)=(-1)^{n}$ with $n=b^{\dagger}b^{\phantom\dagger}$. In the following, we define the projector $\mathcal{P}_{\mathfrak{C}}=\sum_{\eta=\pm} \ket{\mathcal{C}_{\beta_0}^\eta}\bra{\mathcal{C}_{\beta_0}^{\eta}}$ to map our model into the bosonic code limit, where we denote the restricted `cat' Hilbert space by $\mathfrak{C}$ with ground- or steady-state cat amplitudes $\beta=\beta_0$. 

The resonators $b$ and $a_i$ are coupled via non-degenerate three-wave mixing (3WM)~\cite{Walls1983,Heidmann1987,Boyd2009}, 
\begin{equation}\label{eq:3WM}
H_\text{coup}=-g_{3}\left(a_1^{\dagger}b a^{\phantom\dagger}_2+a_2^{\dagger}b^{\dagger}a_1^{\phantom\dagger}\right)\,,
\end{equation}
with amplitude $g_3$. The 3WM coupling can be realised in various ways, e.g., Raman scattering~\cite{Penzkofer1979, Mundhada2019}, optomechanical coupling~\cite{kanari2022two}, and shows up as frequency conversion in nonlinear optical crystals~\cite{Ramelow2009, Coelho2009}, Josephson junctions \cite{Abdo2013}, and polariton condensates~\cite{Carusotto2013, Zambon2019}.  Degenerate and non-degenerate 3WM are related by linear mode coupling~\cite{Frimmer2014, PhysRevLett.128.094301}. 

The 3WM interaction \eqref{eq:3WM} signifies that an excitation from a matter resonator $a_i$ can convert into a pair of excitations that reach another matter resonator $a_j$, $j\neq i$ as well as the intermediate gauge resonator $b$, or vice-versa. Hereby, `hopping' between matter sites $a_i$ is inevitably `recorded' by an excitation arriving or leaving $b$, which imparts a flip in the photon-number parity $\Pi(n)$ in the gauge resonator. 

Using our model~\eqref{eq:full_H}, we show that a $\mathbb{Z}_2$ LGT can be realised by restricting the link resonators to evolve in their respective cat Hilbert spaces [Fig.~\ref{fig:1}(a)]. Projecting Eq.~\eqref{eq:3WM} onto such a subspace brings the projected Hamiltonian $H^{\mathfrak{C}}_\text{coup}$ into the form of a $\mathbb{Z}_2$ light-matter coupling~\cite{Wiese2013}
\begin{align}\label{eq:cat_H}
H^{\mathfrak{C}}_\text{coup}=-\frac{\Omega_R}{2} (a_1^{\dagger}L a_2+a_2^{\dagger}L^{\dagger}a_1)\,,
\end{align}
where an excitation hopping between matter nodes, with Rabi frequency $\Omega_R>0$, is mediated by the `link' (gauge field) operator $L=\mathcal{P}_{\mathfrak{C}}b\mathcal{P}_{\mathfrak{C}}$~\cite{suppmat}. 
Note that $\mathcal{P}_{\mathfrak{C}}$ leaves the matter parts unaffected. 

\begin{figure}
\includegraphics[width=\linewidth]{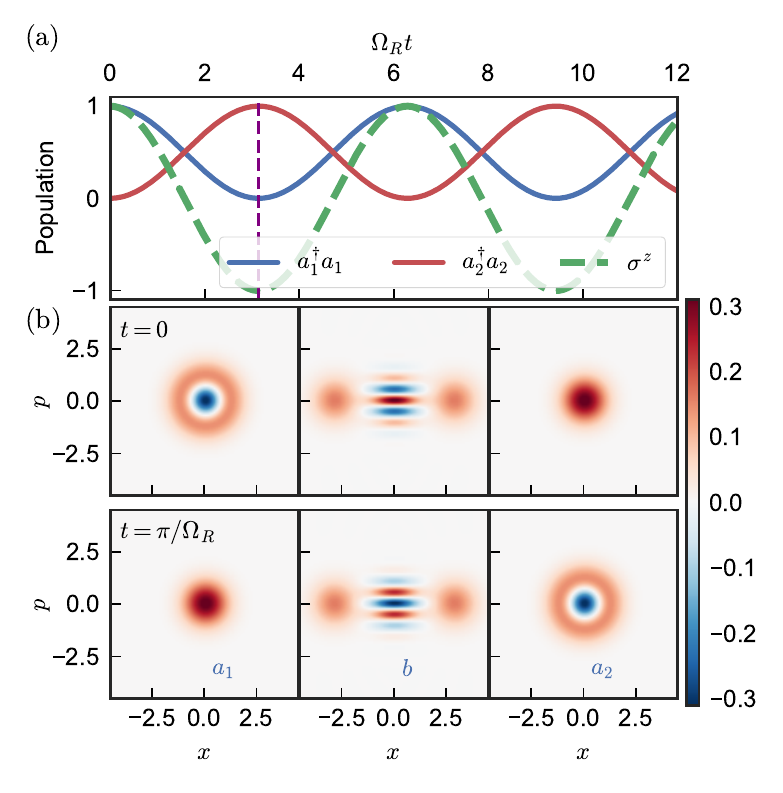}
\caption{(a) Light-matter population dynamics governed by the full Hamiltonian $H$ [cf.~Eq.~\eqref{eq:full_H}] for two matter sites coupled by a link that is realised by a KPO [cf.~Eq.~\eqref{eq:KerrParam}]. The system is initiated in a state with Fock number 1 in resonator $a_1$, and a positive-parity Schrödinger cat state in $b$, denoted $\ket{1,\mathcal{C}_{\beta_0}^+,0}$. The operator $\sigma^z=\ket{\mathcal{C}_{\beta_0}^+}\bra{\mathcal{C}_{\beta_0}^+}-\ket{\mathcal{C}_{\beta_0}^-}\bra{\mathcal{C}_{\beta_0}^-}$ returns the photon number parity for cat states within $\mathfrak{C}$. (b) Marginal Wigner functions for the three resonators at times $t=0$ and $t= \pi/\Omega_R$; the latter corresponds to the vertical dashed line in (a). 
	Here $U=0.03,G=0.12$ ($\beta_0=2$, $\omega_\text{GAP}=0.48$), and $g_3=\omega_{\mathrm{GAP}}/100$.} \label{fig:2}
\end{figure}

Why does our construction entail a $\mathbb{Z}_2$ LGT?
In a $\mathbb{Z}_2$ LGT, (i) the  gauge fields are associated with electric flux operators $E=\sigma^z$ (Pauli matrices $\{\sigma^{x,y,z}\}$), whose values change upon a redistribution of charges with operators $Q_i=(-1)^{a_i^{\dagger}a_i}$. This implies that the electric flux is not conserved, e.g., $L\propto \sigma^x$; (ii) the theory has a $\mathbb{Z}_2$ gauge invariance, namely the Hamiltonian maintains $H=V_iHV_i^{\dagger}$ for local unitary transformations $V_i=e^{i\pi \mathcal{G}_i}$, generated by operators $\mathcal{G}_i=Q_i\sigma^z$. Note that in an extended theory, the local symmetry generators generalise to $\mathcal{G}_i=Q_i\prod_{j:\braket{i,j}}\sigma^z_{i,j}$, where $\sigma^z_{i,j}$ represents the gauge field on the link $\braket{i,j}$ in question. This is also known as the $\mathbb{Z}_2$ analogue of `Gauss law', i.e., the generator is a conserved quantity $[H,\mathcal{G}_i]=0$~\cite{Wiese2013}. 
In our case, when we restrict the Hamiltonian to the cat space, $H_{\mathfrak{C}}=\mathcal{P}_{\mathfrak{C}}H\mathcal{P}_{\mathfrak{C}}$, we (i) recover a similar electric flux operator by also projecting the parity operation $\mathcal{P}_{\mathfrak{C}}\Pi(n+1)\mathcal{P}_{\mathfrak{C}}=\sigma^z$, where the Pauli matrices now act on the cat states $\ket{\mathcal{C}^\pm_{\beta_0}}$~\cite{suppmat}. Moreover, in the large cat amplitude limit, $\beta_0\gg1$, the gauge fields flip the parity, i.e., $L\approx \beta_0 \sigma^x$, as required. Furthermore, property (ii) then holds, leading to the invariance $H_{\mathfrak{C}}=V_iH_{\mathfrak{C}} V_i^{\dagger}$. The $\mathbb{Z}_2$ matter-free dynamics is introduced by a term $E$ in $H^{\mathfrak{C}}_\text{field}$~\cite{schweizer2019floquet,Homeier2021}, inducing rotations $\sim\sigma_{z}$. While cat qubits are robust against such phase flips, several approaches to controllably generate them have been devised, e.g. in Kerr-cat qubits through pulsed, –single and two-photon– drives within $\mathfrak{C}$~\cite{mirrahimi2014dynamically} or involving auxiliary excitations~\cite{Kanao2022}.

Physically, cat states can arise from the competition between two-photon driving (a.k.a.~degenerate 3WM) with Kerr nonlinearity~\cite{Wielinga1993} or using two-photon dissipation~\cite{Gilles1994}. For instance, they manifest as the doubly degenerate ground states of a Kerr parametric oscillator (KPO)~\cite{Puri2019, grimm2019kerr, heugel2019quantum,heugel2022role} [Fig.~\ref{fig:1}(b)]. At a rotating frame at frequency $\omega_b$, the KPO Hamiltonian reads
\begin{align}
\label{eq:KerrParam}
H_\text{field}= -Ub^{\dagger 2}  b^2 + \frac{G}{2}(b^2 + b^{\dagger 2})\,,
\end{align}
with two-photon driving amplitude $G$ (originally at frequency $\approx 2\omega_b$), and Kerr nonlinearity $U$. The spectrum of the KPO~\eqref{eq:KerrParam} is gapped relative to the cat states with splitting approximated by $\omega_\text{GAP}= 4U \beta_0^2$~\cite{Puri2019}, where $\beta_0=\sqrt{G/2U}$ [Fig.~\ref{fig:1}(c)]. The cat eigenfrequencies are degenerate at $U\beta_0^4$, such that the projected Hamiltonian simply reads $H_\text{field}^{\mathfrak{C}}=U\beta_0^4 \mathds{1}_{2\times 2}$, with a trivial zero-point frequency shift.

We turn, now, to show that the effective $\mathbb{Z}_2$ LGT  $H^{\mathfrak{C}}$ well approximates the full dynamics~\eqref{eq:full_H}, see Fig.~\ref{fig:2}. We assume the gap frequency is much larger than the 3WM, $\omega_\text{GAP}\gg \braket{H_\text{coup}}$, and take degenerate matter sites, $\omega_1=\omega_2$. Initially, we place a single excitation in $a_1$, and an even cat in $b$, i.e., $\ket{\psi(t=0)}=\ket{1,\mathcal{C}_{\beta_0}^{+},0}$. We time evolve the state using Eq.~\eqref{eq:full_H} with the KPO model for the gauge sites~\eqref{eq:KerrParam}. We observe coherent population exchange between matter modes, with Rabi frequency $\Omega_R\approx 2g_3\beta_0$, accompanied by cat parity flips. The time snapshots of the resonators' Wigner quasi-probabilities confirm that the Rabi oscillations transition between the states $\ket{1,\mathcal{C}_{\beta_0}^{+},0}$ and $\ket{\psi(t=2\pi/\Omega_R)}=\ket{0,\mathcal{C}_{\beta_0}^{-},1}$ [Fig.~\ref{fig:2}(b)]. 

\begin{figure}
\centering
\includegraphics[width=\linewidth]{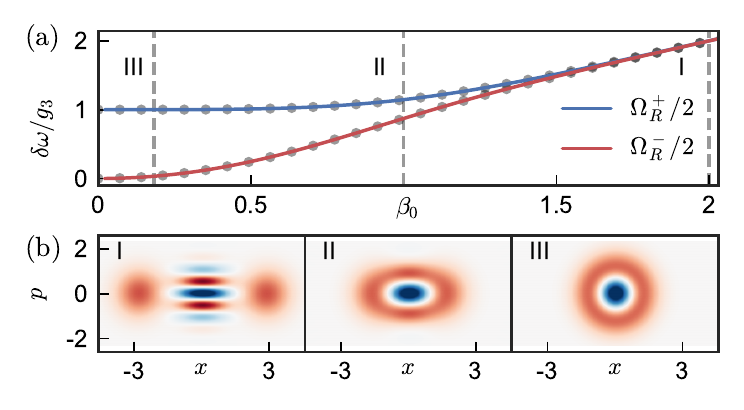}
\caption{(a) Matrix elements $\braket{1,\mathcal{C}^\pm_{\beta_0},0|H_\mathfrak{C}|0,\mathcal{C}^\mp_{\beta_0},1}=\Omega_R^\pm/2$ of the projected Hamiltonian between opposite-parity cat states. The analytical expression for the splitting is superimposed to the numerically exact result $\braket{1,\mathcal{C}^\pm_{\beta_0},0|H|0,\mathcal{C}^\mp_{\beta_0},1}$ (grey dots). Values of $\beta_0$ in (a) are shown as dashed lines. (b) Wigner distribution of the gauge mode $b$, initiated in $\ket{1,\mathcal{C}_{\beta_0}^+,0}$, after a full Rabi swap ($t=\pi/\Omega_R$) with decreasing amplitude $\beta_0=2,1,0.18$ along the columns. In these simulations, we choose $\Omega_R\ll\omega_\text{GAP}$ to prevent excitation outside $\mathfrak{C}$, and take $U=0.03$.}
\label{fig:3}
\end{figure}	

We have shown that  $H$ implements a $\mathbb{Z}_2$ LGT when the cat amplitude $\beta_0$ is large and the 3WM is moderate. What are the limitations of our construction? We first address the impact of lowering $\beta_0$, e.g., by weakening the two-photon drive $G/U\rightarrow 0$. Specifically, we repeat the procedure in Fig.~\ref{fig:2}, and extract the matrix element in $H$ responsible for Rabi oscillations, see Fig.~\ref{fig:3}(a). We observe that, for sufficiently low $\beta_0$, a different Rabi oscillation frequency appears if we start in the odd or the even cat state~\cite{suppmat}.  

This difference marks a discrepancy between the light-matter interaction rates and a corresponding breakdown of the $\mathbb{Z}_2$ LGT dynamics for low $\beta_0$. It crucially originates from cat states shrinking into Fock states, $\ket{\mathcal{C}_{\beta_0}^\pm}\mapsto \{\ket{0}, \ket{1}\}$ as $\beta_0\rightarrow0$ [Fig.~\ref{fig:3}(b)]. This reduction `restores' a $U(1)$ gauge symmetry to our model: (i) as $\beta_0\rightarrow0$, the Wigner functions for $b$ go from mirror-symmetric at all times to full rotationally-symmetric; and (ii) the projectors in $L=\mathcal{P}_\mathfrak{C}b\mathcal{P}_\mathfrak{C}$ approximate the gauge fields to $L\sim \ket{C_\beta^-}\bra{C_\beta^+}$ for $\beta_0\ll1$, and thus the light-matter coupling becomes invariant under arbitrary $U(1)$ transformations ($e^{i\varphi_ i\mathcal{G}_i}$ with $\varphi_i\in (0,2\pi)$)~\cite{suppmat}. Moreover, the shrinking cat subspace implies that the projected states are no longer degenerate, and the time dynamics involve a distinct transition between Fock states. Such asymmetry reflects in the matter-field Rabi frequencies for $|1,\mathcal{C}^+_{\beta_0},0\rangle\leftrightarrow|0,\mathcal{C}^-_{\beta_0},1\rangle$ and $|1,\mathcal{C}^-_{\beta_0},0\rangle\leftrightarrow|0,\mathcal{C}^+_{\beta_0},1\rangle$ processes, which become $\Omega^{\pm}_R=2g_3\beta_0\sqrt{(1\pm e^{-2\beta_0^{2}})/(1\mp e^{-2\beta_0^{2}})}$. 

\begin{figure}
\centering
\includegraphics[width=\linewidth]{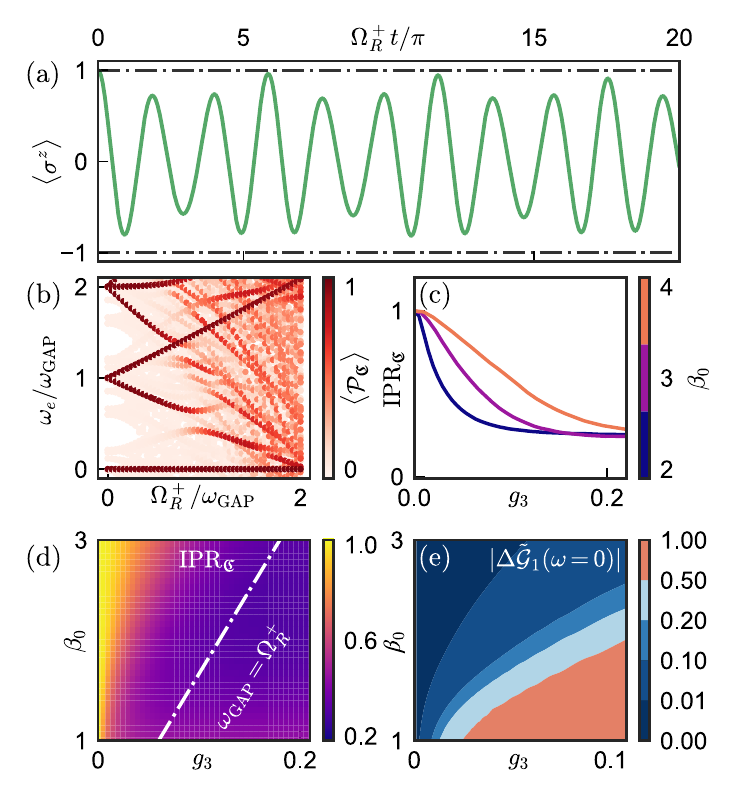}
\caption{(a) Evolution of $\braket{\sigma^z}$ for increasing 3WM. Dashed lines depict the limit where dynamics remains confined in the cat space $\mathfrak{C}$. (b) Eigenspectrum of the full Hamiltonian $H$ for a single link $a_1-b-a_2$. The colorscale marks the value of the projection over the cat subspace. Here and onwards, we assumed the resonance condition $\omega_i=\omega_{\text{GAP}}$. (c) Inverse participation ratio for cat states in the full spectrum, where $\mathrm{IPR}_{\mathfrak{C}}=1$ indicates full localisation in $\mathfrak{C}$. (d) 2D map of $\mathrm{IPR}_{\mathfrak{C}}$ as a function of $\beta_0$ and $g_3$. (e) Normalised DC component of $\Delta\mathcal{G}_i(t)$, marking a baseline deviation from Gauss' law. In these panels, $U=0.03$.}
\label{fig:4}
\end{figure}

We now address the impact of increasing the 3WM. The procedure in Fig.~\ref{fig:2}(a) with  $\braket{H_{\text{coup}}}\sim\omega_{\text{GAP}}$ and $\ket{\psi(t=0)}=\ket{1,\mathcal{C}_{\beta_0}^+,0}$ [Fig.~\ref{fig:4}(a)] reveals the average electric flux beats with a reduced amplitude $\text{max}\braket{\sigma^z}<1$, and additional frequencies. This is in stark contrast with~Figs.~\ref{fig:2} and~\ref{fig:3}, where Rabi exchange at a single frequency $\Omega_R^\pm$ dominates the dynamics. The excitation spectrum of $H$ also reveals an intricate structure [Fig.~\ref{fig:4}(b)], in which cat states and excitations hybridise as $\Omega_R^+>\omega_\text{GAP}$. 
We quantify the overall hybridisation of the cat states $\ket{\mathcal{C}_{\beta_0}^\pm}$ with the system's state $\ket{\psi}$ using the inverse participation ratio $\mathrm{IPR}_{\mathfrak{C}}=\langle\psi|\mathcal{P}_{\mathfrak{C}}|\psi\rangle^2/\langle\psi|\mathcal{P}_{\mathfrak{C}}|\psi\rangle$ [Fig.~\ref{fig:4}(c)]. 
The cat states delocalise over many excited states (low $\mathrm{IPR}_{\mathfrak{C}}$) as the 3WM grows and $\beta_0$ decreases. In fact, the mixing becomes relevant when $\beta_0<g_3/(2U)$, i.e., precisely when $\Omega_R^+>\omega_{\text{GAP}}$, [Fig.~\ref{fig:4}(d)]. Therefore, the hybridisation between $\mathfrak{C}$ and
$\mathfrak{C}^\perp$ subspaces is the cause for the multiple harmonics in Fig~\ref{fig:4}(a)~\cite{suppmat}.

\begin{figure}
\centering
\includegraphics[width=\linewidth]{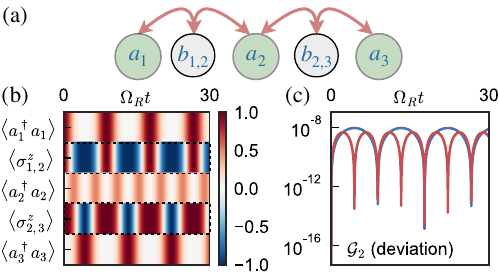}
\caption{(a) A small-scale 1+1 D $\mathbb{Z}_2$ LGT with our  resonator chain scheme. (b) The time evolution of resonators' populations and effective parities on cat subspaces $\mathfrak{C}_i$ starting from a single excitation in $a_1$. The colorscale marks $\braket{a_i^{\dagger}a_i}$ and $\braket{\sigma^z_{i,i+1}}$, respectively. Dashed lines separate between distinct resonator data. (c) Relative deviation of $\braket{\mathcal{G}_2^{\mathrm{chain}}}$ over time. We start from an initial gauge-symmetric state, as in (b). We use, $U=0.03$, $G=0.24$, and $g_3=\omega_\text{GAP}/100$. The numerical calculations are facilitated by the truncation scheme for gauge resonators, as discussed in the text.}
\label{fig:5}
\end{figure}

Strong 3WM hybridisation hampers the performance of even a single $\mathbb{Z}_2$ link. To see the extent of the deviation from an LGT, we explore how much and long the gauge symmetry, expressed by Gauss' laws $[H,\mathcal{G}_i]=0$, is conserved. The evolution must be confined to an eigenspace of the symmetry generators $\mathcal{G}_i$, or equivalently, the value of $\braket{\mathcal{G}_i}$ or `static charge' must be conserved. In the moderate 3WM limit (cf.~Figs.~\ref{fig:2} and~\ref{fig:3}) the dynamics are spanned by states $\ket{1,\mathcal{C}_{\beta_0}^{+},0}$ and $\ket{0,\mathcal{C}_{\beta_0}^{-},1}$, both of which are eigenvectors of $\mathcal{G}_i$ with static charge $-1$. However, for large 3WM amplitudes, the dynamics from $\ket{\psi(t=0)}$ reach eigen-excitations of $H$ with an ill-defined static charge. Similarly, hybridisation causes a breakdown of Gauss' laws and produces errors in the simulation of $\mathbb{Z}_2$ light-matter dynamics. Specifically, the time-evolved operator variances $\Delta \mathcal{G}_i=\braket{\mathcal{G}_i^2}-\braket{\mathcal{G}_i}^2$, which must be zero with an eigenstate of $\mathcal{G}_i$, oscillate around a nonzero baseline. The location of this baseline, obtained from $|\Delta\tilde{\mathcal{G}}_i(\omega=0)|/\text{max}_{J,\beta_0}(|\Delta\tilde{\mathcal{G}}_i(\omega=0)|)$ where $\Delta\tilde{\mathcal{G}}_i(\omega)=\mathrm{FT}[\Delta\mathcal{G}_i(t)]$, gives an estimate of the $\mathbb{Z}_2$ simulation error [Fig.~\ref{fig:4}(e)]. Our results reveal an ideal operation regime (errors below 1$\%$) in the large $\beta_0$ and low $g_3$ region. 

Until now, we  examined the performance of a single bosonic code link as a building block for a $\mathbb{Z}_2$ LGT. Next, we connect two such links in a minimal 1+1 D theory with two gauge fields linking three matter sites [Fig.~\ref{fig:5}(a)]. In Fig.~\ref{fig:5}(b), we study the dynamics initialising the system with a single matter excitation at one end of the chain. The excitation propagates from left to right while flipping the cat parity of the gauge resonators along its path, even when it bounces back off the boundary. Gauge invariance is preserved during the evolution, as indicated by the low values of the central site relative error $|\braket{\mathcal{G}^{\mathrm{chain}}_2(t)}-\braket{\mathcal{G}^{\mathrm{chain}}_2(0)}|/\braket{\mathcal{G}^{\mathrm{chain}}_2(t)}$ in Fig.~\ref{fig:5}(c), where $\mathcal{G}^{\mathrm{chain}}_2=\sigma^z_{1,2}Q_2\sigma^z_{2,3}$. Note that the system's dynamics is intrinsically tied to electric field terms $\sim\sigma^z_{i,i+1}$ and 
the on-site repulsion on the matter sites $\sim(a_i^{\dagger}a_i)^2$, which we neglected here. The latter interaction terms could play a crucial role in initial state preparation, an area for future research to explore.

In the analysis of Fig.~\ref{fig:5}, we rely on truncation to the lowest $M$ eigenstates of $H_\mathrm{field}$ instead of employing a Fock basis for $b$, which would require much larger truncated Hilbert spaces to properly describe cat states and their dynamics~\cite{suppmat}. Since the model conserves the number of matter excitations, this reduces the dimension of the computational Hilbert space, $\mathcal{H}$, for $N$ matter sites and $N-1$ gauge resonators to $\mathrm{dim}(\mathcal{H})=NM^{N-1}$. This method provides an accurate description of the dynamics of a chain of links. Note, however, that the exponential growth of $\mathrm{dim}(\mathcal{H})$ still implies that a proper description of larger systems requires the assistance of quantum simulation devices. 

Our proposal motivates the realisation of LGT quantum simulation using bosonic code networks. In light of contemporary results in the field~\cite{grimm2019kerr}, the realisation of the basic building block of the model is within experimental reach. Note, however, that with superconducting devices, future work will need to address the interplay between the out-of-equilibrium drives and dissipation for a more realistic description. Nevertheless, with the abundance of new bosonic code platforms available, the Hamiltonian description presented here is sufficiently broad to inspire LGT emulation in other systems. The natural redundancy in bosonic Hilbert spaces and the spectral isolation of cats guarantee the robustness of our proposal and opens the door to studying improvements of the simulations using error-correction schemes~\cite{mirrahimi2014dynamically, grimm2019kerr}. Exploring systems with higher dimensionality would require additional Hamiltonian interactions~\cite{suppmat}. Furthermore, our study sets the foundation to 1+1 D LGTs~\cite{Yao2020} and more involved symmetries, such as $SU(2)$ LGTs~\cite{Zohar2013} by incorporating additional bosonic codes. 

We thank A.~Eichler for critical reading of the manuscript and T.~L.~Heugel and C.~Schweizer for fruitful discussions. J.d.P. was supported by the ETH Fellowship program (grant no. 20-2 FEL-66). O.Z.  acknowledges funding through SNSF Sinergia grant CRSII5\_198631 and from the Deutsche Forschungsgemeinschaft (DFG) - project number 449653034.

\bibliography{refs}

	\clearpage
	\pagebreak
	\renewcommand{\theequation}{S\arabic{equation}}
	\renewcommand{\thefigure}{S\arabic{figure}}
	\setcounter{equation}{0}
	\setcounter{figure}{0}
	\setcounter{table}{0}
	\setcounter{page}{1}
	
	\onecolumngrid
	\begin{center}
		\textbf{\large Supplemental Material: Dynamical gauge fields with bosonic codes}
	\end{center}
	\setcounter{secnumdepth}{2} 
	\twocolumngrid

		\section{Lattice gauge theories in a nutshell}
	In this section, we briefly introduce the fundamentals of Lattice Gauge Theories (LGT) as a discretisation limit of dynamical gauge field theories (DGFT). We focus on the simplest case of quantum electrodynamics. Here, we do not aim to provide a comprehensive review of the literature on the subject but refer the interested reader to Ref.~\cite{Wiese2013_,jendrzejewski2017dynamical_}, which uses Hamiltonian language and Refs.~\cite{wiese2009introduction_,Gattringer2010_, Wipf2013_} which are more focused on a path integral formulation.
	
	We focus on Quantum Electrodynamics (QED), a DGFT that explains how electronic and positronic fields interact via photonic fields, the latter being the `gauge bosons' of the theory. In this theory, gauge bosons are non-interacting, in contrast with more complex DGFTs such as quantum chromodynamic theories, where gauge bosons or `gluons' directly interact with each other. The Lagrangian density for QED reads
	\begin{equation}\label{eq:lag_density}
		\mathcal{L}=\bar{\psi}(i\gamma^{\mu}D_\mu-m)\psi-\frac{1}{4}F^{\mu,\nu}F_{\mu,\nu}\,.
	\end{equation}
	Here, $\psi(x)$ and $\bar{\psi}(x)=\psi(x)^{\dagger}\gamma^0$ are 4-component Dirac spinors describing electrons and positrons with mass $m$, while $\gamma^{\mu}$ are  Dirac matrices. The covariant derivative $D_\mu=\partial_\mu+ieA_\mu(x)$ ($\mu=\{1,2,3,4\}$) contains the 4-vector potential $A_\mu(x)=[\Phi(x),-\textbf{A}(x)]$, whose field strength tensor $F_{\mu,\nu}=\partial_\mu A_\nu(x)-\partial_{\nu}A_\mu(x)$ desribes the electric and magnetic fields, $\textbf{E}(x)$ and $\textbf{B}(x)$, respectively~\cite{Jackson1999_}. The Lagrangian density~\eqref{eq:lag_density} is invariant under (Abelian) $U(1)$ transformations
	\begin{align}\label{eq:gauge_trafo}
		\psi(x)\mapsto& e^{ie\alpha(x)}\psi(x),&&\bar{\psi}(x)\mapsto\bar{\psi}e^{-ie\alpha(x)},\\
		A_{\mu}(x)\mapsto& A_\mu(x)-\partial_{\mu}\alpha(x)\,.
	\end{align}
	
	\subsection{Discretisation into a Wilson LGT}
	Quantum electrodynamics and other DGFTs suffer from ultraviolet divergences, which can be removed by regularisation and subsequent renormalisation. At the perturbative level, this is accomplished by regularising individual Feynman diagrams. To apply regularisation in non-perturbative regimes, one can apply Wilson regularisation~\cite{Wilson1974_}. This entails the discretisation of continuous space-time into a hyper-cubic grid with dimension $D$ and lattice spacing $a$, which imposes an ultraviolet momentum cutoff $1/a$. The coupling constants in the theory then become a function of $a$ and are ultimately tuned to reach the continuum limit $a\rightarrow0$. The naive discretisation of the Dirac equation leads to the fermion doubling problem; for each physical fermion, one finds $2^d-1$ unphysical species~\cite{Wipf2013_}. Although solutions to this problem exist through the use of, e.g. `domain wall fermions'~\cite{Kaplan1992_}, the additional fermionic degrees of freedom can then be reduced and partly reinterpreted as fermion `flavours'. An example of this idea is in the model of `staggered' or Kogut–Susskind fermions~\cite{Kogut1975_}. In this model, fermionic fields $\psi_x,\psi_y$, (with anticommutators $\{\psi_x^{\dagger},\psi_y\}=\delta_{x,y}$, $\{\psi_x,\psi_y\}=\{\psi_x^{\dagger},\psi_y^{\dagger}\}=0$) are spin-less, but the positions in the lattice map to the spins in the continuum theory. Defining the hopping amplitude $t$ and focusing on the model without external fields, the Hamiltonian reads~\cite{Wiese2013_}
	\begin{equation}\label{eq:stag}
		H_{\mathrm{free}}=-t\sum_{\braket{xy}}s_{x,y}(\psi_x^{\dagger}\psi_y+\psi_y^{\dagger}\psi_x) + m\sum_{x}s_x\psi_x^{\dagger}\psi_x\,.
	\end{equation}
	Here, hopping occurs between sites $x$ and $y=x+\hat{k}$, with $\hat{k}$ being a vector of length $a$ in the spatial $k$-direction. The sign factors $s_{x}$ and $s_{x,y}$ play the role of the fermionic $\gamma$-matrices. Namely,  $s_x=(-1)^{x_1+\cdots+x_d}$ is associated with nodes in the lattice, while the sign factors for the links in the $1^{\rm st}$-direction read $s_{x,y} = 1$, for those in the  $2^{\rm nd}$-direction $s_{x,y}=(-1)^{x_1}$, and for the links in the $k^{\rm th}$-direction  $s_{x,y} = (-1)^{x_1+\cdots+x_{k-1}}$. In effect,  the alternating sign of the mass term implies that the model simultaneously incorporates matter and antimatter fields by using a single degree of freedom on bipartite lattices. 
	
	At this stage, we introduce a background electromagnetic field with vector potential $A(x)$. On a lattice, the vector potential $A_k(x)$ along dimension $k$ is associated with a directed link in the lattice, $x\rightarrow y$. In the continuum limit, the LGT Hamiltonian must remain $U(1)$ invariant to respect~\eqref{eq:gauge_trafo}. To accomplish this, we introduce the unitary parallel transporter operator $U_{x,y}=e^{ie\int_{x_k}^{x_k+a}dx_k\hspace{1mm}A_k(x)}\in U(1)$, which implements the Peierls substitution similar to that of a single charged particle with charge $e$, where momentum $\textbf{p}\rightarrow \textbf{p}-e \textbf{A}(x)$~\cite{jendrzejewski2017dynamical_}
	\begin{equation}\label{eq:stag_QED_static}
		H_\text{coup}=-t\sum_{\braket{xy}}s_{x,y}(\psi_x^{\dagger}U_{x,y}\psi_y+\psi_y^{\dagger}U_{x,y}^{\dagger}\psi_x) + m\sum_{x}s_x\psi_x^{\dagger}\psi_x\,.
	\end{equation}
	The vector potential operator on a link $A_{x,y}$ can be derived from $U_{x,y}=e^{iaeA_{x,y}}$. The electric field operator associated with a link then reads $E_{x,y}=-i\partial/\partial(aeA_{x,y})$ with electron charge $e$,  which recovers the classical limit $\mathbf{E} = -\nabla \phi(\mathbf{r},t) + \partial_t \mathbf{A}(\mathbf{r},t)$. These definitions induce the commutation relations $[E_{x,y},U_{x'y'}]=\delta_{x,x'}\delta_{y,y'}U_{x,y}$ and $[E_{x,y},U^{\dagger}_{x'y'}]=-\delta_{x,x'}\delta_{y,y'}U_{x,y}^{\dagger}$. In the lattice representation, magnetic fields are associated with plaquette products $U_\square=U_{w,x}U_{x,y}U_{z,y}^{\dagger}U^{\dagger}_{w,z}$. Adding the free electromagnetic field dynamics ($H_\text{field}=\int dx\hspace{1mm}\frac{1}{2}(\textbf{E}^2+\textbf{B}^2)$) upgrades Eq.~\eqref{eq:stag_QED_static}  to
	\begin{align}\label{eq:stag_QED}
		H_\mathrm{QED}=&-t\sum_{\braket{x,y}}s_{x,y}(\psi_x^{\dagger}U_{x,y}\psi_y+\psi_y^{\dagger}U_{x,y}^{\dagger}\psi_x) +\\
		& m\sum_{x}s_x\psi_x^{\dagger}\psi_x + \frac{e^2}{2}\sum_{\braket{x,y}}E_{x,y}^2-\frac{1}{4e^2}\sum_\square(U_\square+U_\square^{\dagger})\,,\nonumber
	\end{align}
	where the $\square$ symbol regroups indexes for a square plaquette.
	
	\subsection{Gauge invariance and Gauss' law}
	In classical electromagnetism, Gauss' law establishes a relation between the divergence of the electric field $\nabla\cdot\textbf{E}$ (mapping  into $\sum_k (E_{x,x+\hat{k}}-E_{x-\hat{k},x})$ in an LGT) and the charge density $\rho$ (proportional to $-\psi_{x}^{\dagger}\psi_x$). However, in Hamiltonian formulation, this can not be implemented as an operator identity since  $\nabla \cdot \textbf{E}\neq0$ in the absence of charges~\cite{Wiese2013_}, and the scalar potential does not show up in~Eq.~\eqref{eq:stag_QED}. Nevertheless, gauge symmetry implies a similar (local) conservation law that combines the matter and the electric field. For this conservation, we can define a generator of infinitesimal gauge transformation at site $x$ that takes the form
	\begin{equation}
		\mathcal{G}_x^{\mathrm{QED}}=\psi_x^{\dagger}\psi_x+\sum_k(E_{x,x+\hat{k}}-E_{x-\hat{k},x})\,,
	\end{equation}
	such that $[H_\mathrm{QED},\mathcal{G}^{\mathrm{QED}}_x]=0$. A general gauge transformation $V=\Pi_x e^{ie\alpha(x)\mathcal{G}^{\mathrm{QED}}_x}$ is then generated by $\mathcal{G}^{\mathrm{QED}}_x$. It acts as $
	V\psi_xV^{\dagger}=\Omega_x\psi_x,$ and $VU_{x,y}V^{\dagger}=\Omega_xU_{x,y}\Omega_y^{\dagger}$, with $\Omega_x=e^{ie\alpha(x)}$. Gauss' law implies that the physical Hilbert space is divided into sectors of eigenvalues of $\mathcal{G}^{\mathrm{QED}}_x$ or `static charges', i.e., $\mathcal{H}=\oplus_{g_x}\mathcal{H}({g_x})$. Each such sector $\mathcal{H}({g_x})$ has eigenvectors $\ket{\psi({g_x})}\in \mathcal{H}({g_x})$ with well defined static charge $\mathcal{G}^{\mathrm{QED}}_x\ket{\psi({g_x})}=g_x\ket{\psi({g_x})}$. The only gauge-invariant states of the physical Hilbert space can be defined as those for which the Gauss law is fulfilled, e.g., $\mathcal{G}^{\mathrm{QED}}_x\ket{\psi}=0$~\footnote{In a non-Abelian group, defining the charges is more involved ~\cite{Zohar2013_}}.

	\subsection{Quantum link model and discrete gauge theories}
	The Hilbert space of a Wilson LGT is infinite-dimensional, even for a single link. Each link variable, the parallel transporter $U_{x,y}$, is analogous to a  particle moving on the `group manifold',  which is a circle for the gauge group $U(1)$. Simulating such a continuous degree of freedom is generally a difficult task. Instead, \textit{quantum link models}~\cite{chandrasekharan1997quantum_} are considered to approximate the dynamics, where $U_{x,y}$ are replaced by a discrete degree of freedom while exactly implementing continuous gauge symmetries. An identification which fulfils the electromagnetic commutator relations employs angular momentum operators
	\begin{align}
		U_{x,y}=J_{x,y}^+,&&U_{x,y}^{\dagger}=J_{x,y}^-,&&E_{x,y}=J_{x,y}^z\,,
	\end{align}
	provided $[J_{x,y}^{a},J_{x',y'}^b]=i\delta_{x,x'}\delta_{y,y'}\epsilon_{a,b,c}J_{x,y}^c$, with $a,b,c$ labelling $x,y,z$ coordinates. The reduction to a discrete model comes at the expense of relaxing some commutation relations: while in Wilson's theory $[U_{x,y},U^{\dagger}_{x',y'}]=0$, in a $U(1)$ quantum link model $[U_{x,y}, U_{x',y'}^{\dagger}]=2\delta_{x,x'}\delta_{y,y'}E_{x,y}$.
	
	Even when one restricts to the smallest spin value $1/2$ in this $U(1)$ quantum link model, interesting gauge theories emerge. For example, suppose Gauss's law is appropriately modified. In that case, the $U(1)$ quantum link model turns into a quantum dimer model~\cite{Banerjee2014_}, which is used in condensed matter physics to model systems related to high-temperature superconductors. A $\mathbb{Z}_2$ LGT, such as Kitaev's toric code~\cite{Kitaev2003_}, is similar to such models. The main difference is that the symmetry of the theory gets reduced to that of local $\mathbb{Z}_2$ transformations, namely $V_i=e^{i\pi \mathcal{G}_i}.$
	\begin{align}
		\psi_i\mapsto V_i \psi_iV_i^{\dagger},&&L_{i,i+1}\mapsto V_i L_{i,i+1} V_i^{\dagger}\,,
	\end{align}
	with $\mathcal{G}_i=Q_i\prod_{j:\braket{i,j}}\sigma^z_{i,j}$, where $\sigma^z_{i,j}$ represents the gauge field on the link $\braket{i,j}$ in question and $Q_i=(-1)^{\psi_i^{\dagger}\psi_i}$.
	
	Note that the realisation of an LGT in the main text assumes bosonic matter along 1D $\psi_x\rightarrow a_x$. The lattice does not contain loops or plaquette terms associated with magnetic fields.

	\section{Summary of bosonic error correction}
	Building a fault-tolerant quantum computer with controlled hardware overhead is one of modern physics' great pursued challenges.
	The quick degradation of quantum states due to interaction with the environment is an especially pressing problem when requiring multiple quantum operations in a quantum simulator. One possible strategy to mitigate this source of errors is to employ quantum error correction (QEC). This section provides an overview of the efforts of bosonic QEC. This subject has been exposed in detail in multiple papers and reviews~\cite{nielsen_chuang_2010_, Devoret2013_, RevModPhys.93.025005_,Cai2021_}, so we only summarise it here.
	
	One of the simplest instances of classical error correction is classical repetition codes, which use odd multiples of 0's and 1's to represent logical 0 and 1, respectively. These codes can correct bit-flip errors by using majority voting and effectively suppress a significant number of errors. By doing so, even if some of the information in the encoded message is damaged by noise, there is still enough redundancy in the encoded information to recover the original information fully. QEC works in an analogue manner but through the expansion of the Hilbert space of a logical qubit to create the code space, i.e., the subspace where quantum information can be stored as a superpositions of states (code words). The larger Hilbert space is endowed with local symmetries to add redundancy to the encoded quantum information. A typical approach to this is via multiple entangled qubits that store quantum information non-locally. For QEC to work, the action of noise must lead to predictable errors, transferring quantum information from the logical subspace to another, `error' subspace, orthogonal to it. Before applying appropriate recovery operations, errors must also be able to be detected without perturbing the encoded information. This is typically attained by introducing and measuring ancillary qubits or modes interacting with the logical qubits.
	
	Qubit-based QEC has experienced a surge in activity over the last few years. Most proposals of qubit-based QEC are limited to either bit-flip or phase errors~\cite{Moussa2011_, Schindler2011_, Chen2021_}. A particularly promising avenue is that of surface codes, which exploit the topological properties of qubits arranged in a two-dimensional grid~\cite{Fowler2012_,Chen2021_,Krinner2022_}.
	Nevertheless, these methods are challenging to implement due to the need for sufficiently low error rates -the maximum is 1$\%$ for surface codes and, more critically, a significant resource overhead. A major challenge is scaling the number of qubits and achieving long lifetimes in current platforms (prominent examples are trapped ions, nitrogen-vacancy centres in diamond, and superconducting circuits). Additionally, multi-qubit encoding systems often struggle with logical operations and extended lifetimes due to the increased number of error channels and the need for non-local gates on collections of physical qubits. 
	
	Bosonic QEC tackles the hardware overhead scaling challenge by encoding an intrinsically error-protected qubit into the energy levels of a harmonic oscillator and drawing on the notion of non-locality in oscillator phase space~\cite{mirrahimi2014dynamically_}. A single bosonic mode can provide an infinitely large Hilbert space. By leveraging the redundancy of its Hilbert space, QEC can be implemented in a single degree of freedom by only extending excitation numbers while keeping the noise channels fixed~\cite{Chuang1997_}. The implementation of bosonic QEC involves using a bosonic element, such as a harmonic oscillator, in conjunction with a nonlinear element, typically a two-level qubit, to control and manipulate quantum states. This architecture differs from traditional qubit-based QEC, where quantum information is stored on qubits, and harmonic oscillators are utilised for coupling or readout.
	
	State-of-the-art bosonic codes, including Gottesman-Kitaev-Preskill, cat, and binomial codes, have been subject to extensive research and experimentation and continue to drive progress in quantum information processing~\cite{RevModPhys.93.025005_,Cai2021_}. Cat codes are based on superpositions of coherent states. Coherent states can be readily generated using classical optics and are widely used in communication due to their ability to encode information in their phase, which is less susceptible to photon loss errors. One simple realisation is given by the four-component cat code~\cite{Leghtas2013_}, where quantum information is encoded in a superposition of coherent states with four different phases, which is non-local in the phase space of the harmonic oscillator. This code provides the necessary redundancy to counteract photon loss errors by utilising two states for encoding and two for error detection. Measurements of the photon number parity allow detecting errors, which can be readily realized in a quantum non-demolition manner in a superconducting qubit architecture~\cite{Sun2013_}. 
	
	A simpler realisation, the two-component cat code, encodes information in the phase of a coherent state,  using two coherent states with opposite phases, namely $\ket{\pm\beta}$, as the basis states for the code. This code lacks redundancy and is not error correctable, as a single-photon-loss error will result in the state remaining within the same code space. However, it allows for realisation of robust `cat-qubits', see e.g. experimental realisations in nonlinear microwave cavities~\cite{Vlastakis2013_, Puri2019_, grimm2019kerr_}. Cat qubit realisations stabilise the Hilbert space of nonlinear oscillators into the two-dimensional cat subspace $\mathfrak{C}$. The distance between the two basis states (see, e.g., the Wigner functions in main text Fig. 2) protects against any noise process that induces local displacement in phase space (since $\braket{-\beta|b|\beta}=\beta e^{-2|\beta|^2}$), offering a built-in stabilising mechanism. As a result, phase-flip errors are exponentially suppressed with the average number of bosons, while bit-flip errors only increase linearly, as noted in the reference~\cite{grimm2019kerr_}. Although the Hilbert space of such a cat qubit is not large enough to fully protect it against excitation-loss errors, the extreme difference in rates between bit-flip and phase-flip errors (noise biasing) provides a mechanism for stabilising the encoded information. The distinction between different types of errors can be leveraged to minimise the number of qubits required for QEC since error-correcting codes can be designed specifically to correct the most frequent type of error, in this case, phase flips, rather than allocating resources to correct both amplitude and phase errors.
	
	\stepcounter{enumi}
	\setcounter{equation}{0}
	
	\begin{figure}[h]
		\includegraphics[width=\linewidth]{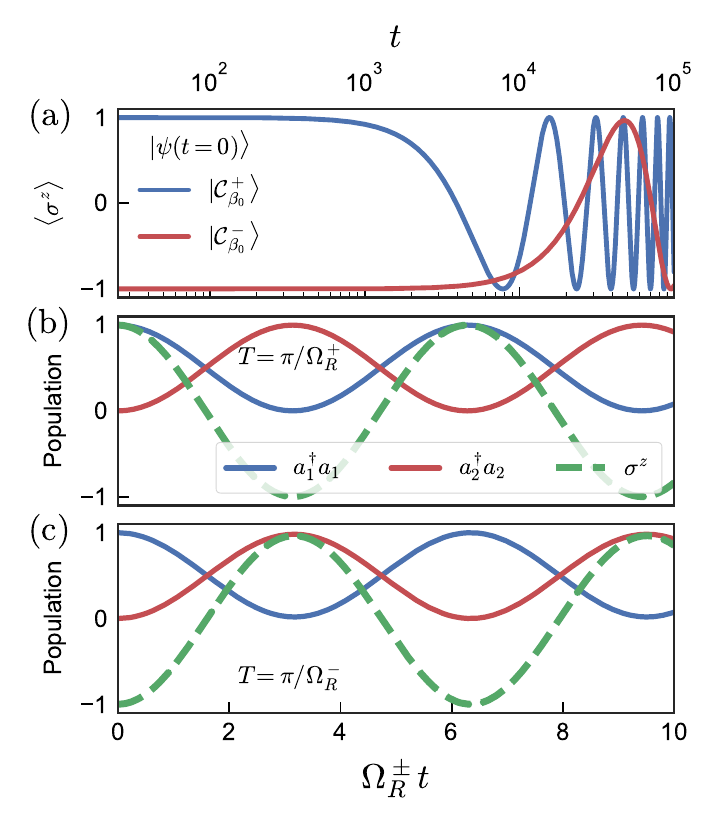}
		\caption{(a) Time evolution of $\braket{\sigma^z}$ in a link $a_1-b-a_2$, with a $b$ resonator governed by $H_\text{field}$ [main text Eq.~(5)] for $\beta_0=0.4$, starting from $\ket{\mathcal{C}_{\beta_0}^+}$ or $\ket{\mathcal{C}_{\beta_0}^-}$. (b-c) Dynamics with re-scaled frequencies $\Omega_R^\pm$, respectively. Other parameters are $U=0.03$.} \label{fig:time_evo_diff}
	\end{figure}
	
	\section{Projection over cat subspace}\label{sec:projection}
	In the main text, we demonstrate that the projection of the resonator lattice model can emulate an LGT in the bosonic code limit. Here, we provide additional details of the involved algebraic manipulations. 
	
	We define the projector over the cat subspace of the gauge resonator $b$ as $\mathcal{P}_{\mathfrak{C}}=\sum_{\eta=\pm} \ket{\mathcal{C}_{\beta_0}^\eta}\bra{\mathcal{C}_{\beta_0}^{\eta}}$, to project our model to the bosonic code limit. Keeping the notation in the main text, the restricted `cat' Hilbert space at each link is dubbed by $\mathfrak{C}$. In order to project $H_\text{coup}$ [cf.~Eq.~(3) in main text] onto the cat space, it is important to note that (i) the coherent states $\ket{\beta}$ are eigenstates of the annihilation operator $b$, namely $b\ket{\beta}=\beta\ket{\beta}$, but (ii) not of the creation operator: $b^{\dagger}=\beta\ket{\beta}+D(\beta)\ket{1}$, where $D(\beta)=e^{\beta b^{\dagger}-\beta^*b}$ denotes a displacement operator and $\ket{1}$ the Fock state with one excitation. We can use these relations to  obtain $L=\mathcal{P}_{\mathfrak{C}}b\mathcal{P}_{\mathfrak{C}}$ in closed form. We find
	\begin{align}
		L=& u^x(\beta_0)\sigma^x+iu^y(\beta_0)\sigma^y,\label{eq:b_cat}\\
		L^{\dagger}\approx& u^x(\beta_0)\sigma^x-iu^y(\beta_0)\sigma^y\label{eq:bd_cat}.
	\end{align}
	Here, we defined the Pauli basis $\{\sigma^{x,y,z}\}$ acting onto the Schrödinger cat states of resonator $b$. In this expression, we observe that the link operators in the effective LGT [cf.~Eq.~(4) in the main text] contain Hermitian ($\propto \sigma^x$) and anti-Hermitian contributions ($\propto i \sigma^y$), weighted by cat-size-dependent factors $u^{x,y}(\beta_0)=\beta_0(\mathcal{N}_{\beta_0}^{-}/\mathcal{N}_{\beta_0}^{+}\pm\mathcal{N}_{\beta_0}^{+}/\mathcal{N}_{\beta_0}^{-})/2$, with $\mathcal{N}_{\beta_0}^{\pm}=1/\sqrt{2(1\pm e^{-2\beta_0^2})}$. These factors arise from the non-orthogonality of coherent states.
	
	\subsection{Vanishing cat-amplitude limit}
	As detailed in the main text, for large cat amplitudes $L\sim\beta_0\sigma^x$. This results from the limit $u_{x}(\beta_0)\simeq \beta_0$, $u_y(\beta_0)\simeq 0$ when $\beta_0\gg 1$, which cancels the above mentioned anti-Hermitian contribution. An inspection of Eq.~(4) in the main text then shows that since the $\sigma^x$ operator acting on the cat basis is defined as $\sigma^x=\ket{\mathcal{C}_{\beta_0}^+}\bra{\mathcal{C}_{\beta_0}^-}+\ket{\mathcal{C}_{\beta_0}^-}\bra{\mathcal{C}_{\beta_0}^+}$, the light-matter Rabi frequencies become irrespective of the cat parity. On the other hand, we illustrate in the main text that a decreasing cat amplitude results in a $U(1)$-symmetric LGT with Rabi frequencies that depend on the starting condition. In fact, in the limit $\beta_0\rightarrow0$, we have $u^{x,y}(\beta_0)\rightarrow 1$, i.e. $L\approx\ket{\mathcal{C}_{\beta_0}^-}\bra{\mathcal{C}_{\beta_0}^+}$. Matter-field dynamics initialised in $\ket{\mathcal{C}_{\beta_0}^+}$ will have a Rabi frequency equal to $\Omega_R^+=2g_3$, while starting from $\ket{\mathcal{C}_{\beta_0}^-}$ will imply no evolution at all, since $\Omega_R^-=0$. The differences in Rabi frequency for arbitrary $\beta_0$ follow from the fact the two matrix elements
	\begin{equation}\label{eq:matrix_el}
		|\braket{\mathcal{C}_{\beta_0}^\mp|L|\mathcal{C}_{\beta_0}^\pm}|=u_x(\beta_0)\pm u_y(\beta_0)\,,
	\end{equation}
	are different. To support the data in main text Fig.~3, we show the time evolution of the system initialised in either $\ket{1,\mathcal{C}_{\beta_0}^+,0}$ or $\ket{1,\mathcal{C}_{\beta_0}^-,0}$ (order $a_1-b-a_2$, as in main text) in the low $\beta_0$ limit in Fig.~\ref{fig:time_evo_diff}. In Fig.~\ref{fig:time_evo_diff}(a)  a disparate oscillation period appears for both initial conditions, while in Figs.~\ref{fig:time_evo_diff}(b) and (c) we demonstrate that such period is $T^\pm=\pi/\Omega_R^\pm$; exactly matching the prediction of Eq.~\ref{eq:matrix_el}.

	\begin{figure}[t!]
		\includegraphics[width=\linewidth]{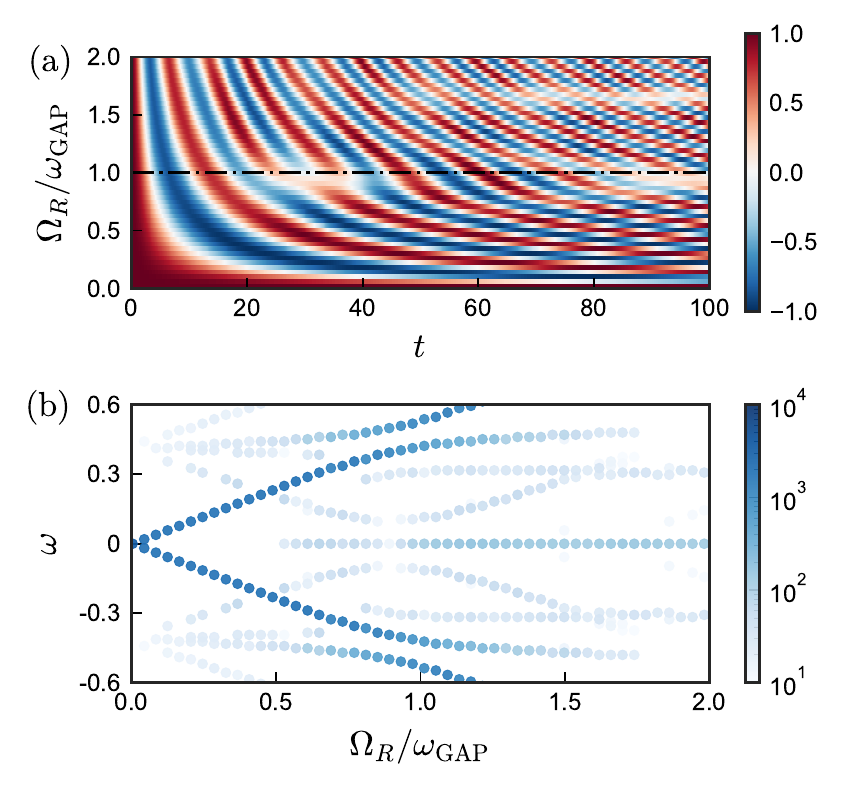}
		\caption{(a) Time evolution of $\braket{\sigma^z}$ in a $a_1-b-a_2$ link, governed by $H$ for increasing 3WM. The dashed line marks the region where Rabi frequencies for light-matter interaction and the gap frequency between $\mathfrak{C}$ and $\mathfrak{C}^\perp$ become identical. (b) Fourier frequencies for increasing 3WM, extracted from $\braket{\tilde{\sigma}^z(\omega)}=\text{FT}[\braket{\sigma^z(t)}]$ The colour scale represent their amplitudes in arbitrary units. In these panels $U=0.03$ and $\beta_0=2$, such that $\Omega_R^+\approx\Omega_R^-\equiv\Omega_R$.} \label{fig:sigma_z_mix}
	\end{figure}
	
	\subsection{Hybridisation for strong 3WM}
	The approximate symbol in Eq.~\eqref{eq:bd_cat} precisely stands for the fact that the action of $b$ leaks excitations away from $\mathfrak{C}$. This leakage is responsible for the mixing of ground (cat) states and excitations in $b$ and leads to the breakdown of gauge invariance in the regime where $\Omega_R^\pm\sim\omega_{\text{GAP}}$ [cf.~main text Fig.~4]. We observe that one of the manifestations of this effect is the reduction of the amplitude of oscillations in $\braket{\sigma^z(t)}$ [cf.~Fig.~4(a)]. Here, we extend this analysis by depicting the time evolution in Fig.~\ref{fig:sigma_z_mix} for multiple values of the 3WM. We see a single-oscillatory period that decreases with $g_3$, up to the point where $\Omega_R$ becomes of a similar order as $\omega_\text{GAP}$: Dynamics then become strongly perturbed [Fig.~\ref{fig:sigma_z_mix}(a)] and multimode, i.e. with many relevant Fourier components [Fig.~\ref{fig:sigma_z_mix}(b)].
	
	Excitation leakage is also evidenced by the explicit action of $b^{\dagger}$ over cat states, namely $b^{\dagger}|\mathcal{C}_{\beta_0}^{\pm}\rangle =|\mathcal{C}_{\beta_0}^{\pm}\rangle+|\mathcal{E}_{\beta_0}^{\mp}\rangle$, where the states $|\mathcal{E}_{\beta_0}^{\pm}\rangle=\mathcal{N}_{\beta_0}^{\pm}\left(|1,\beta_0\rangle\mp|1,-\beta_0\rangle\right)$ are superpositions of displaced Fock states $D(\beta_0)|n\rangle=|n,\beta_0\rangle$. Note that the number parity $\Pi(n)$ of the states $|\mathcal{E}_{\beta_0}^{\pm}\rangle$ is opposite to the relative phase between the states in the superposition. Through 3WM, cat states thus couple to excitations with opposite parity, with leakage rate given by first-order perturbation theory using Fermi's Golden rule~\cite{cohen1986quantum_} $\Gamma_{\text{leak}}\propto|\langle\mathcal{E}_{\beta_0}^{\mp}|b^{\dagger}|\mathcal{C}_{\beta_0}^{\pm}\rangle|^{2}$. The relevant matrix element follows straightforwardly
	\begin{align}
		\langle\mathcal{E}_{\beta_0}^{\mp}|b^{\dagger}|\mathcal{C}_{\beta_0}^{\pm}\rangle=	
		\pm\mathcal{N}_{\beta_0}^{\mp}\mathcal{N}_{\beta_0}^{\pm}\left(\langle1|D(-2\beta_0)|0\rangle+\langle1|D(2\beta_0)|0\rangle\right),
	\end{align}
	where we employed the relations $D(\beta_{1})D(\beta_{2})=D(\beta_{1}+\beta_{2})$ and $D^{\dagger}(\beta)=D(-\beta)$. The right-hand side of the last term contains an overlap between displaced harmonic oscillators $F_{n,m}(\alpha,\beta)=\langle n|D(\alpha)D^{\dagger}(\beta)|m\rangle$, and in particular the terms $F_{n,m}(\mp\beta_0,\pm\beta_0)$. The calculation is identical to the so-called Franck-Condon factors in molecular physics~\cite{May2003_}. The factors $F_{n,m}$ are known to decay exponentially with $\beta_0^{2}$ as $F_{n,m}(-\beta_0,\beta_0)\sim f(\beta_0^2) e^{-2\beta_0^{2}}$ where $f(\beta^2)$ is a polynomial function of $\beta^2$. This implies that when $\beta_0\gg1$, then $\langle\mathcal{E}_{\beta_0}^{\mp}|b|\mathcal{C}_{\beta_0}^{\pm}\rangle\approx0$. Note that this limit is equivalent for the KPO to have  $\omega_{\text{GAP}}\gg\Omega_{R}^{\pm}$ for fixed $U,g_{3}$.

	The above derivation can be readily extended to multiple links. This requires a projector that collapses all link resonators $b_{i,j}$, corresponding to links $i-j$, to their respective bosonic code limits. This is given by $\mathcal{P}_\mathfrak{C}=\bigotimes_{\braket{i,j}} \mathcal{P}_{\mathfrak{C}_{i,j}}$, with projectors  $\mathcal{P}_{\mathfrak{C}_{i,j}}=\sum_{\eta=\pm} \ket{\mathcal{C}_{\beta_{i,j}}^\eta}\bra{\mathcal{C}_{\beta_{i,j}}^{\eta}}$ onto the cat subspace sitting on each link $\braket{i,j}$, $\mathfrak{C}_{i,j}$, with population $\beta_{i,j}$. In this way, the link operators generalise to $L_{i,j}= u^x(\beta_{i,j})\sigma_{i,j}^x+iu^y(\beta_{i,j})\sigma_{i,j}^y$ and $L_{i,j}^{\dagger}\approx u^x(\beta_{i,j})\sigma_{i,j}^x-iu^y(\beta_{i,j})\sigma_{i,j}^y$, with link Pauli matrices $\{\sigma_{i,i+1}^{x,y,z}\}$.
	
	\begin{widetext}
		\section{Gauge invariance}
		\stepcounter{enumi}
		\setcounter{equation}{0}
		In the main text, we introduce the effective $\mathbb{Z}_2$ gauge symmetry the system can experience in the bosonic code limit. In this section, we show the explicit proof of $\mathbb{Z}_2$ gauge invariance for the projected single link Hamiltonian $H^{\mathfrak{C}}$ [cf.~Eq.~(4) in the main text] with $L=\beta_0\sigma^x$ and gauge symmetry generator $\mathcal{G}_i=(-1)^{a_i^{\dagger}a_i}\sigma^z$ ($i,j=\{1,2\}$ labels matter sites). The proof of $[H^{\mathfrak{C}},\mathcal{G}_i]=0$ is straightforward by noting that  $ Q_i a_i = -a_iQ_i$, $\sigma^x\sigma^z=-\sigma^z\sigma^x$, since $Q_i$ is the photon parity operator and the Pauli matrices anticommute, namely $\{\sigma^x,\sigma^z\}=0$.
		Therefore,
		\begin{align}
			[H_{\text{coup}}^{\mathfrak{C}},\mathcal{G}_{i}]=&	-\frac{\Omega_{R}}{2}\beta_0\left[[a_{j}^{\dagger}a_{k}+a_{k}^{\dagger}a_{j},Q_{i}]\sigma^{x}\sigma^{z}+Q_{i}\left(a_{j}^{\dagger}a_{k}+a_{k}^{\dagger}a_{j}\right)[\sigma^{x},\sigma^{z}]\right]=\nonumber\\
			&-\frac{\Omega_{R}}{2}\beta_0\left(-2Q_{i}\delta_{i,j}\left(a_{j}^{\dagger}a_{k}+a_{k}^{\dagger}a_{j}\right)-2Q_{i}\delta_{i,k}\left(a_{j}^{\dagger}a_{k}+a_{k}^{\dagger}a_{j}\right)+2Q_{i}\left(a_{j}^{\dagger}a_{k}+a_{j}^{\dagger}a_{k}\right)\right)\sigma^{x}\sigma^{z}=0\,.
		\end{align}
		for either $i=j$ or $i=k$. The demonstration proceeds identically in the generic case of Eqs.~\eqref{eq:b_cat} and \eqref{eq:bd_cat}, since  $\sigma^x\sigma^y=-\sigma^y\sigma^x$. Note also the proof can be readily generalised to systems with multiple links by employing a symmetry generator that accounts for all the links $k$ connected to a matter site $i$, i.e. $\mathcal{G}_i=(-1)^{a_i^{\dagger}a_i}\prod_{j:\braket{i,j}} \sigma^z_{i,j}$.

		\section{Excitations on top of cat subspace: numerical assessment}
		
		Here, we provide further details about the numerical time evolution and eigenstate calculations reported in the main text, Figs.~4,5. In numerical simulations, the Hilbert space of a single $a_1-b-a_2$ link, namely $\mathcal{H}_{a_1-b-a_2}$, is truncated. For instance, in Fock space, one can keep a Hilbert's space dimension equal to $\text{dim}(\mathcal{H}_{a_1-b-a_2})=N_\mathrm{matter}^2N_\mathrm{gauge}$ states, where $N_\mathrm{matter},N_\mathrm{gauge}$ are Fock cutoffs for the matter and gauge resonators, respectively. Due to the strong nonlinearity $U$ and two-photon drive $G$ in resonator $b$, the eigenstates of $H_\text{field}$ contain multiple excitations in the number basis. Hence, $N_\mathrm{gauge}$ needs to be sufficient to simulate the cat state features faithfully (e.g. their Poissonian number distribution~\cite{Wielinga1993_,Cochrane1999_}) and their excitations. 
		
		\begin{figure}[ht!]
			\includegraphics[width=0.7\linewidth]{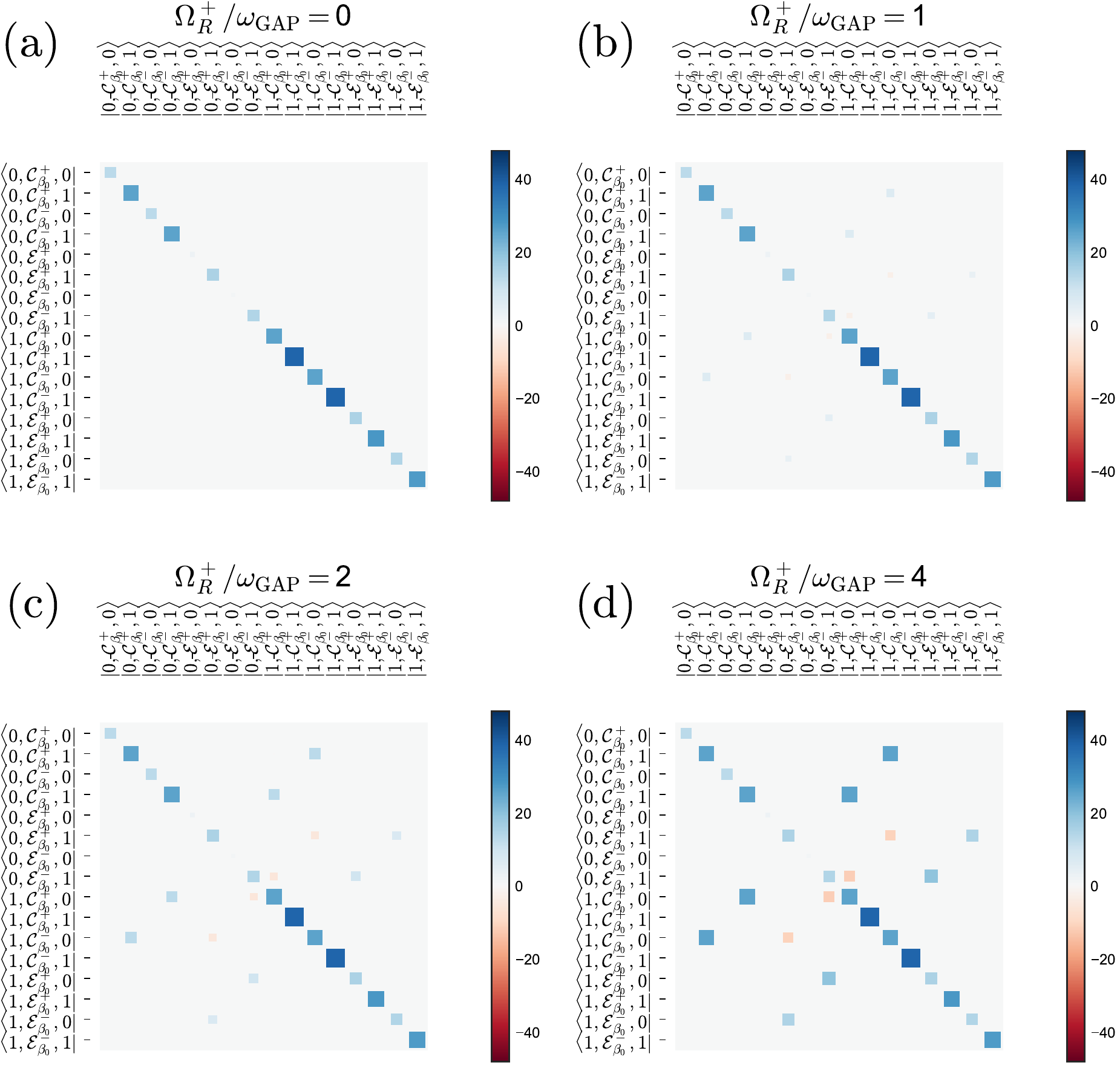}
			\caption{Hinton diagrams for the Hamiltonian matrix $H$ of a link $a_1-b-a_2$, where the states in $b$ are expressed in the eigenbasis of $H_\text{field}$ [main text Eq. (5)]. The elements of such a Hamiltonian are real. Blue and red squares represent positive and negative values, and the size of each square represents the magnitude of each value. Panels (a-d) are calculated with increasing values of 3WM, with a fixed $U=1.$ and cat amplitude $\beta_0=2$. The calculation includes the cat states $\ket{C_{\beta_0}^\pm}$ in $b$ and the first excited states dubbed as $\ket{\mathcal{E}_{\beta_0}^\pm}$, obtained by pre-diagonalising $H_\mathrm{field}$ in the Fock basis with $N_\mathrm{gauge}=100$.} \label{fig:hinton}
		\end{figure}
		
		The parameters employed in main text Figs.~1,2, and 3 (moderate 3WM) imply that it is often sufficient to keep an intermediate cutoff $N_\mathrm{gauge}<30$ and $N_\mathrm{mat}<10$. In fact, the model approximately preserves the number of matter resonator excitations (main text Eq. (4)), implying that, in fact is enough to consider $N=2$ excited states in the matter. We can further understand why such approximation holds and reveal the action of 3WM by keeping the Fock basis for matter sites but rotating into the eigenbasis of $H_\mathrm{field}$ for $b$ sites. A diagonal Hamiltonian in the absence of 3WM [Fig.~\ref{fig:hinton}(a)] develops off-diagonal contributions as the 3WM increases. Namely, the action of the 3WM in this regime is to induce couplings between cat-like states mainly, as the Hamiltonian structure for $\beta_0=2$ reveals [compare Figs.~\ref{fig:hinton}(a) and (b)]. Here we employ the numerically diagonalised $H_\text{field}$. Still, note we have reflected in the labels that the excitations of $H_\text{field}$ are well approximated by displaced Fock states $\ket{\mathcal{E}_{\beta_0}^\pm}$ defined in Sec.~\ref{sec:projection}. This approximation neglects tunnelling between the two displaced harmonic oscillator states sitting at either side of the effective system's potential [cf. main text Fig. 1(c)], i.e. $F_{n,n}\rightarrow0$, a valid assumption provided $\beta_0=2$ in this calculation.
		
		The effective confinement of dynamics within $\mathfrak{C}$ for moderate 3WM leads to the approximate fulfilment of Gauss' law, as described in the main text. As 3WM increases, the coupling between states with weight in $\mathfrak{C}$, namely $\ket{\mathcal{C}_{\beta_0}^\pm}$ and the excited states in $\mathfrak{C}^{\perp}$, i.e., $\ket{\mathcal{E}_{\beta_0}^\pm}$, becomes comparable to the inter-cat-state coupling [Figs.~\ref{fig:hinton}(c) and (d)]]. This is the case for the values chosen in simulations for main text Fig. 4. Convergence can require up to $N_\mathrm{gauge}=100$ and $N_\mathrm{mat}=30$ Fock states, i.e. $\text{dim}(\mathcal{H}^{\text{eff}}_{a_1-b-a_2})\sim 10^4$. Simulations are still feasible for a single or few links, but scaling of the required Hilbert space is exponential. To alleviate this `combinatorial explosion', we note low values of off-diagonal matrix elements between $\ket{\mathcal{C}_\beta^\pm}$ and  $\ket{\mathcal{E}_\beta^\pm}$ relative to the inter-cat matrix elements. This implies cat states are still an appropriate basis for a moderate 3WM of $\Omega_R^+\sim\omega_\text{GAP}$ and the number of required eigenstates for $b$, $M$, is significantly lower than the Fock cutoff ($M\ll N_\mathrm{gauge}$).
	\end{widetext} 
	
	\section{Route towards plaquette terms}
	
	\begin{figure}[ht]
		\includegraphics[width=\linewidth]{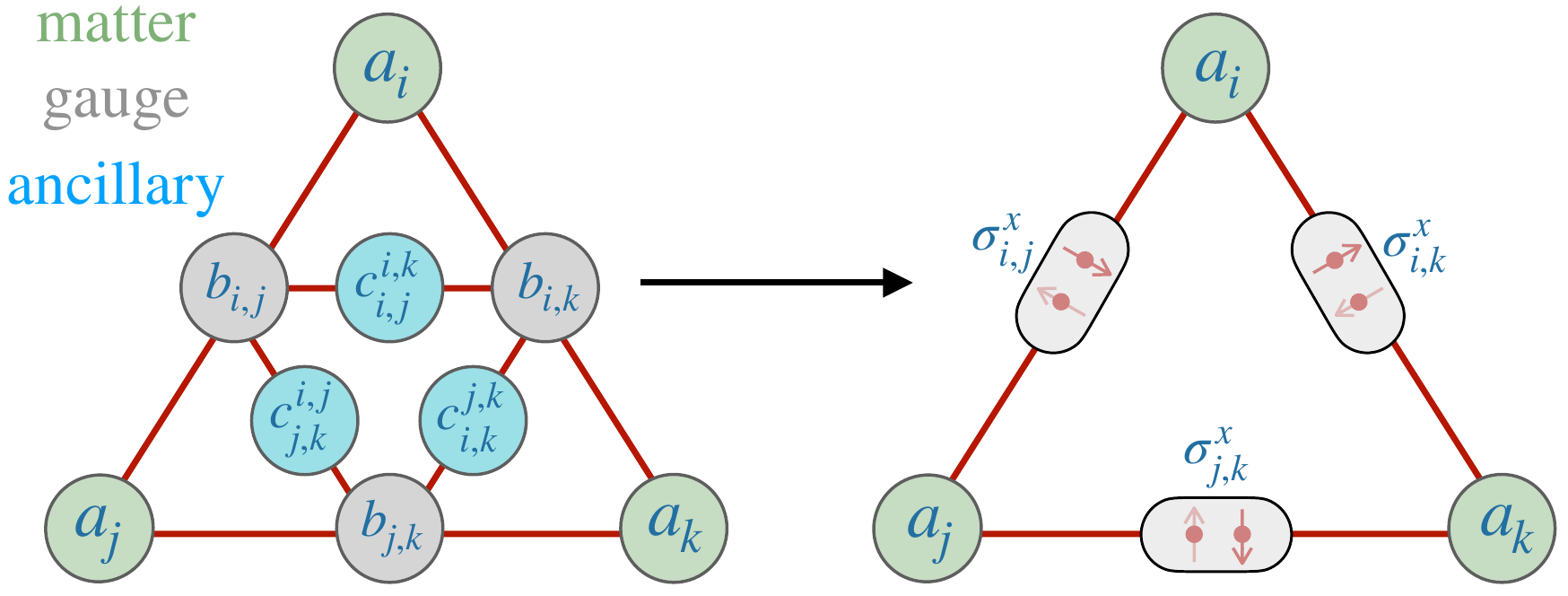}
		\caption{Implementation of plaquette terms in a $\mathbb{Z}_2$ LGT in a triangular geometry. Assuming the bosonic code limit for gauge resonators (grey) and restricting the number of excitations in the ancillary resonators (blue) brings plaquette terms between effective $\mathbb{Z}_2$ gauge fields.} \label{fig:plaquette}
	\end{figure}

	The extension of DGFTs to 2+1 D requires the inclusion of plaquette terms in the Hamiltonian [as per Eq.\ref{eq:stag_QED}].
	In a $\mathbb{Z}_2$ theory, these terms consist of products of gauge operators for the links composing a plaquette. The required interactions that need to be isolated depend on the geometry of the lattice; for instance, they are required to be 4-body interactions for the toric code in a square plaquette. To incorporate these terms, one possible approach is to use non-degenerate multi-wave mixing interactions~\cite{Sameti2016_}, e.g. $H_{\text{coup}}^{\text{plaquette}}=g_3\sum_{\langle i,j,k\rangle:\triangle}b_i^{\dagger}b_j b_k+\mathrm{H.c.}$ in a triangular plaquette. In the bosonic code limit and for $\beta_0\gg 1$, this interaction approximates to $H_{\text{coup}}^{\text{plaquette}}\approx 2g_3\sum_{\langle i,j,k\rangle:\triangle}\beta_i\beta_j\beta_k\hspace{1mm} \sigma^x_i\sigma^x_j\sigma^x_k $, where the symbol $\triangle$ denotes the plaquette indexes. 
	
	Another approach that works for generic $\mathbb{Z}_2$ LGTs making use of ancillary sites is inspired by Ref.~\cite{Homeier2021_}. Consider a triangular plaquette with three matter sites ($a_i$) connected to three gauge resonators ($b_{i,j}$) as shown in Fig.~\ref{fig:plaquette}. The matter-gauge dynamics are described by a generalisation of Eq.~3, given as $H_{\text{coup}}=-g_3\sum_{i,j:\triangle}(a_i^{\dagger}b_{i,j}a_j+a_j^{\dagger}b_{i,j}^{\dagger}a_i)$. Each gauge resonator is coupled to two ancillary resonators (annihilation operators $c_{i,j}^{j,k}$). Here sub-indexes denote initial and final gauge resonator indexes in counter-clockwise order. The effective interaction Hamiltonian has the form
	\begin{equation}
		H_{\text{ancillary}}^{\triangle}= -g_3\sum_{\langle i,j,k,l\rangle:\small\downtriangle}(b_{i,j}^{\dagger}c_{i,j}^{k,l}b_{k,l}+b_{k,l}^{\dagger}(c_{i,j}^{k,l})^{\dagger}b_{i,j}),
	\end{equation}
	where $\downtriangle$ refers to the indices in the upside-down inner plaquette of gauge resonators in Fig.~\ref{fig:plaquette}. In the bosonic code limit for large $\beta_0$, this Hamiltonian becomes:
	\begin{equation}\label{eq:ancilla_H_BC}
		H_{\text{ancillary}}^{\triangle}= -\sum_{\langle i,j,k,l\rangle:\downtriangle}g_3^{i,j,k,l}(\sigma^x_{i,j}c_{i,j}^{k,l}\sigma^x_{k,l}+\sigma^x_{k,l}(c_{i,j}^{k,l})^{\dagger}\sigma^x_{i,j}),
	\end{equation}
	with effective couplings $g_3^{i,j,k,l}=g_3\beta_i\beta_j\beta_k\beta_l$.
	
	In the Hamiltonian (\ref{eq:ancilla_H_BC}), link operators couple with one another with couplings mediated by the ancillary oscillators. In the following, we assume the effective couplings are equal $g_3\beta_i\beta_j\beta_k\beta_l\equiv g_\triangle$. The inner plaquette is self-dual, so the Hamiltonian can also be simplified by reducing the number of redundant indexes in the ancillary resonators. This results in the following simplified form:
	\begin{equation}
		H_{\text{ancillary}}^{\triangle}= -g_{\triangle}\sum_{i,j:\triangle}(c_i^\dagger\sigma^x_{i,j}c_j+c_j^\dagger\sigma^x_{i,j}c_i),
	\end{equation}
	which is equivalent to the Hamiltonian in~\cite{Homeier2021_}, except for the quantisation axis in the Pauli matrices. 
	
	This model can be understood as a hopping model with operator-valued hopping phases that depend on the link operators as $\varphi_{i,j}=\pi\sigma^x_{i,j}/2$. To simplify the Hamiltonian, we perform a phase shift transformation on the resonators $c_i$ using the operator-dependent phase shift $Uc_jU^{\dagger}=e^{-i\sum_j\theta_j}c_j$, where $\theta_j=\pi(\sigma_{j,j+1}^x-\sigma^x_{j-1,j})/2$. This evenly distributes the dynamical hopping phases. Next, we express the ancillary modes in the momentum eigenbasis using $\tilde{c}_k=\sum_{j=1}^3 e^{-i2\pi kj/3}c_j/\sqrt{3}$ for $k=\{-1,0,1\}$, resulting in
	\begin{equation}\label{eq:triag_plaquette}
		H_{\text{ancillary}}^{\triangle}=-2g_{\triangle}\sum_{k=\{-1,0,1\}}\cos\big(\frac{2\pi k +\Phi}{3}\big)\tilde{c}_k^{\dagger}\tilde{c}_k,
	\end{equation}
	where $\Phi$ is a link-operator-dependent flux. 
	
	In the Hamiltonian (\ref{eq:triag_plaquette}), as $\Phi$ changes from $0$ to $\pi$, the Hamiltonian simulates a plaquette term. To see this, we must note that (i) The operator valued flux is $\Phi\sim\Pi_{j:\triangle}\sigma_{j,j+1}^{x}$ and (ii) the low energy subspace is gapped, with a gap directly proportional to $\Phi$. With regards to (i): Phase accumulation is, in fact, equivalent to transforming the wavefunction of a matter excitation via $e^{i\sum_{i,j:\triangle}\varphi_{i,j}}=(e^{i\pi(\sigma_{12}^x-\sigma_{31}^x)/2})\sigma_{1,2}^xe^{i\pi(\sigma_{23}^x-\sigma_{12}^x)/2})\sim \prod_{i:\triangle}\sigma_{i,i+1}^x$ if the phases are evenly distributed. Similarly, since $\Pi_{j:\triangle}e^{i\pi\sigma_{j,j+1}^{x}/2}=\Pi_{j:\triangle}\sigma_{j,j+1}^{x}$, phase accumulation in a $\mathbb{Z}_2$ equates to having a flux $\Phi\sim\Pi_{j:\triangle}\sigma_{j,j+1}^{x}$. This observation allows us to address (ii): If $\Phi=0$, the lowest eigenfrequency is $-2g_\triangle$ (corresponding to the mode $k=0$) whereas if $\Phi=\pi$, is a degenerate subspace with eigenfrequency $-g_\triangle$ (formed by, e.g., the $k=0$ and $k=-1$ modes).  If the dynamics is restricted to this low energy manifold gapped by $g_\triangle$, then the Hamiltonian (\ref{eq:triag_plaquette})  simulates a plaquette term: $H_{\text{coup}}^{\text{plaquette}}\mapsto -g_\triangle(\Phi-\pi)\sim \Pi_{j:\triangle}\sigma_{j,j+1}^{x}$. 
	
	By following a similar process as described in Ref.~\cite{Homeier2021_}, it is possible to extend this theoretical construction to assemble multi-plaquette systems and lattices.

\end{document}